# Singular Mueller matrices

**JOSÉ J. GIL,**[1*] **RAZVIGOR OSSIKOVSKI,**[2] **IGNACIO SAN JOSÉ**[3]

[1]*Universidad de Zaragoza. Pedro Cerbuna 12, 50009 Zaragoza, Spain*
[2]*LPICM, CNRS, Ecole Polytechnique, Université Paris – Saclay, 91128 Palaiseau, France*
[3]*Instituto Aragonés de Estadística. Gobierno de Aragón. Bernardino Ramazzini 5, 50015 Zaragoza, Spain*
*\*Corresponding author: ppgil@unizar.es*

Singular Mueller matrices play an important role in polarization algebra and have peculiar properties that stem from the fact that either the medium exhibits maximum diattenuation and/or polarizance, or because its associated canonical depolarizer has the property of fully randomizing the circular component (at least) of the states of polarization of light incident on it. The formal reasons for which the Mueller matrix $\mathbf{M}$ of a given medium is singular are systematically investigated, analyzed and interpreted in the framework of the serial decompositions and the characteristic ellipsoids of $\mathbf{M}$. The analysis allows for a general classification and geometric representation of singular Mueller matrices, of potential usefulness to experimentalists dealing with such media.

*OCIS codes: (260.2130) Ellipsometry and polarimetry; (260.5430) Polarization; (290.5855) Scattering, polarization.*

## 1. INTRODUCTION

In spite of the possible difficulties in determining, in practice, if a measured Mueller matrix can be considered as being singular or not because of the limited precision of the experimental data, singular Mueller matrices play an important role in Mueller algebra, as well as in the understanding of the polarimetric behavior of certain kinds of media, and thus deserve a specific study.

Obviously, experimentally determined Mueller matrices are affected by errors, and in accordance with the precision of the polarimetric device, it is necessary to establish appropriate criteria for deciding if a measured Mueller matrix $\mathbf{M}_{\exp}$ is considered singular in practice. That is, when the value of $\det \mathbf{M}_{\exp} \approx \Delta$, where $\Delta$ is the predetermined tolerance, $\mathbf{M}_{\exp}$ can be submitted to the filtering process indicated in Section 3, in such a manner that the filtered Mueller matrix $\mathbf{M}_f$ satisfies exactly the property $\det \mathbf{M}_f = 0$.

A family of media whose associated Mueller matrices are singular is constituted by polarizers and analyzers (linear, circular, or elliptic in general), which are quite common in experimental arrangements. Polarizers are provided by optical industry in the form of polarizing film sheets (dichroic polarizers), crystalline components (Glan-Taylor, Glan-Thompson, Rochon…), wire grid elements, etc. with very efficient extinction ratio, so that their representation by means of singular Mueller matrices is usually justified.

Moreover, polarizers and analyzers, characterized respectively by unit polarizance and unit diattenuation are not the only devices with a singular Mueller matrix; specific media with depolarizing behavior are also characterized by Mueller matrices with zero determinant.

Therefore, the aim of this work is to formally determine, parameterize and geometrically represent the polarimetric properties of media whose associated Mueller matrices are singular. To do so, we consider first, in Sec. 2, several approaches for the serial decomposition of a given Mueller matrix $\mathbf{M}$, which provide the appropriate framework for identifying certain classes of components that are characterized by singular Mueller matrices. Section 3 deals with the geometric representation of $\mathbf{M}$ through a set of characteristic ellipsoids that adopt particular degenerate forms when $\mathbf{M}$ is singular. Then Section 4 is devoted to the analysis, classification and representation of the different kinds of singular Mueller matrices.

To simplify some mathematical expressions, let us bring out the partitioned block expression of a Mueller matrix [1],

$$\mathbf{M} = m_{00} \begin{pmatrix} 1 & \mathbf{D}^T \\ \mathbf{P} & \mathbf{m} \end{pmatrix}$$

$$\mathbf{D} \equiv \frac{1}{m_{00}} (m_{01}, m_{02}, m_{03})^T$$

$$\mathbf{P} \equiv \frac{1}{m_{00}} (m_{10}, m_{20}, m_{30})^T$$

$$\mathbf{m} \equiv \frac{1}{m_{00}} \begin{pmatrix} m_{11} & m_{12} & m_{13} \\ m_{21} & m_{22} & m_{23} \\ m_{31} & m_{32} & m_{33} \end{pmatrix}$$

(1)

where the **D** and **P** are respectively called the *diattenuation vector* and the *polarizance vector* of $\mathbf{M}$ [2]. The absolute values of these vectors are called *diattenuation*, $D \equiv |\mathbf{D}|$ and *polarizance*, $P \equiv |\mathbf{P}|$.

Both polarizance $P$ and diattenuation $D$ have dual nature depending on the direction of propagation of light (forward or reverse) [3,4]; $D$ is both the diattenuation of $\mathbf{M}$ and the polarizance of the *reverse Mueller matrix* $\mathbf{M}^r \equiv \mathrm{diag}(1,1,-1,1) \mathbf{M}^T \mathrm{diag}(1,1,-1,1)$ [5,6] corresponding to the same interaction as $\mathbf{M}$, but interchanging the input and output directions ($\mathbf{M}^T$ being the transposed matrix of $\mathbf{M}$). Conversely, $P$ is both the polarizance of $\mathbf{M}$ and the diattenuation of $\mathbf{M}^r$.

A medium with $P = 1$ and $D < 1$ is called a *depolarizing polarizer*; a medium with $D = 1$ and $P < 1$ is called a *depolarizing analyzer* and a medium satisfying $D = P = 1$ is called a *polarizer* (also *nondepolarizing polarizer* or *polarizer-analyzer*) [2].

The *degree of polarimetric purity* [3] of $\mathbf{M}$ is given by the *depolarization index* [7]

$$P_\Delta = \sqrt{D^2 + P^2 + \|\mathbf{m}\|_2^2}\big/\sqrt{3} \tag{2}$$

where $\|\mathbf{m}\|_2$ stands for the Frobenius norm of the submatrix $\mathbf{m}$ of $\mathbf{M}$. Further, the *mean intensity coefficient* of $\mathbf{M}$ (i.e., transmittance or reflectance for unpolarized input states) is given by $m_{00}$.

Mueller matrices associated with systems that do not depolarize any totally polarized input state for both forward and reverse directions (i.e., whose depolarization index satisfies $P_\Delta = 1$) are called *pure Mueller matrices* (also *nondepolarizing* or *Mueller-Jones matrices*), while Mueller matrices satisfying $P_\Delta < 1$ are called *nonpure* or *depolarizing* Mueller matrices.

## 2. SERIAL DECOMPOSITIONS OF SINGULAR MUELLER MATRICES

A serial decomposition of a Mueller matrix $\mathbf{M}$ consists in representing $\mathbf{M}$ as a product of particular Mueller matrices. Its physical meaning is that the whole system is considered as a pile of polarization components disposed sequentially along the pathway of the probing electromagnetic wave. This arrangement of the components constitutes the *serial equivalent system*. Prior to summarizing such approaches for the serial decomposition of a given Mueller matrix $\mathbf{M}$ that will be useful for the study of singular Mueller matrices, let us recall that, in general, both the polar decomposition and the singular value decomposition of $\mathbf{M}$ lead to unphysical components [3], and thus, other decompositions ensuring their own physical realizability have to be considered.

### A. Generalized polar decomposition

The generalized polar decomposition (also called Lu-Chipman decomposition or forward decomposition) is formulated as [2]

$$\mathbf{M} \equiv m_{00} \begin{pmatrix} 1 & \mathbf{D}^T \\ \mathbf{P} & \mathbf{m} \end{pmatrix} = m_{00} \hat{\mathbf{M}}_{\Delta P} \mathbf{M}_R \hat{\mathbf{M}}_D$$

$$\hat{\mathbf{M}}_{\Delta P} \equiv \begin{pmatrix} 1 & \mathbf{0}^T \\ \mathbf{P}_{\Delta P} & \mathbf{m}_{\Delta P} \end{pmatrix}, \mathbf{M}_R \equiv \begin{pmatrix} 1 & \mathbf{0}^T \\ \mathbf{0} & \mathbf{m}_R \end{pmatrix}, \hat{\mathbf{M}}_D \equiv \begin{pmatrix} 1 & \mathbf{D}^T \\ \mathbf{D} & \mathbf{m}_D \end{pmatrix},$$

$$\mathbf{P}_{\Delta P} = \frac{\mathbf{P} - \mathbf{m}\mathbf{D}}{1 - D^2}, \mathbf{m}_R^{-1} = \mathbf{m}_R^T \ (\det \mathbf{m}_R = +1),$$

$$\mathbf{m}_D = \sqrt{1 - D^2}\, \mathbf{I}_3 + \left(1 - \sqrt{1 - D^2}\right)\hat{\mathbf{D}} \otimes \hat{\mathbf{D}}^T,$$

$$\mathbf{I}_3 \equiv \mathrm{diag}(1,1,1), \quad \hat{\mathbf{D}} \equiv \mathbf{D}/D, \tag{3}$$

where the normalized symmetric matrix $\mathbf{M}_D$ represents a diattenuator whose diattenuation-polarizance vector $\mathbf{D}$ is equal to the diattenuation vector of $\mathbf{M}$; the orthogonal matrix $\mathbf{M}_R$ represents a retarder, and the normalized depolarizer matrix $\hat{\mathbf{M}}_{\Delta P}$, with $\mathbf{m}_{\Delta P} = \mathbf{m}_{\Delta P}^T$, represents a depolarizer with nonzero polarizance and zero diattenuation.

When $\mathbf{M}$ is singular, it necessarily satisfies one of the following possibilities: $\hat{\mathbf{M}}_{\Delta P}$ is singular; $\mathbf{M}_D$ is singular, or both $\hat{\mathbf{M}}_{\Delta P}$ and $\mathbf{M}_D$ are singular.

Observe that the decomposition (**3**) can also be expressed as [8]

$$\mathbf{M} = m_{00}\,\hat{\mathbf{M}}_{\Delta P}\,\mathbf{M}_R\,\hat{\mathbf{M}}_D$$
$$= m_{00}\,\hat{\mathbf{M}}_{\Delta P}\,\hat{\mathbf{M}}'_D\,\mathbf{M}_R = m_{00}\,\mathbf{M}_R\,\hat{\mathbf{M}}'_{\Delta P}\,\hat{\mathbf{M}}_D,$$
$$\left(\hat{\mathbf{M}}'_D \equiv \mathbf{M}_R\,\hat{\mathbf{M}}_D\,\mathbf{M}_R^T,\ \hat{\mathbf{M}}'_{\Delta P} \equiv \mathbf{M}_R^T\,\hat{\mathbf{M}}_{\Delta P}\,\mathbf{M}_R\right). \tag{4}$$

The decomposition (**3**), when applied to the Mueller matrix $\mathbf{M}^T$ (which, as we have seen above, is closely related to the reverse Mueller matrix $\mathbf{M}^r$), adopts the form

$$\mathbf{M}^T \equiv m_{00} \begin{pmatrix} 1 & \mathbf{P}^T \\ \mathbf{D} & \mathbf{m}^T \end{pmatrix} = m_{00}\,\hat{\mathbf{M}}_{\Delta D}^T\,\mathbf{M}_R^T\,\hat{\mathbf{M}}_P \tag{5}$$

where $\hat{\mathbf{M}}_{\Delta D}^T$ is a depolarizer without diattenuation, $\mathbf{M}_R^T$ is the transposed matrix of $\mathbf{M}_R$ (i.e., $\mathbf{M}_R^T$ represents a retarder), and $\hat{\mathbf{M}}_P$ represents a diattenuator whose diattenuation-polarizance vector is equal to the polarizance vector $\mathbf{P}$ of $\mathbf{M}$. Therefore, a *reverse decomposition* of $\mathbf{M}$ is obtained by taking the transpose in Eq. (**5**) [9] (recall that $\hat{\mathbf{M}}_P = \hat{\mathbf{M}}_P^T$)

$$\mathbf{M} = m_{00}\,\hat{\mathbf{M}}_P\,\mathbf{M}_R\,\hat{\mathbf{M}}_{\Delta D};$$

$$\hat{\mathbf{M}}_{\Delta D} \equiv \begin{pmatrix} 1 & \mathbf{D}_{\Delta D}^T \\ \mathbf{0} & \mathbf{m}_{\Delta D} \end{pmatrix},\ \mathbf{M}_R \equiv \begin{pmatrix} 1 & \mathbf{0}^T \\ \mathbf{0} & \mathbf{m}_R \end{pmatrix},\ \hat{\mathbf{M}}_P \equiv \begin{pmatrix} 1 & \mathbf{P}^T \\ \mathbf{P} & \mathbf{m}_P \end{pmatrix} \tag{6}$$

where the normalized depolarizer $\hat{\mathbf{M}}_{\Delta D}$ exhibits nonzero diattenuation and zero polarizance, while the retarder $\mathbf{M}_R$ coincides with that of the forward decomposition.

As with the forward decomposition, the reverse decomposition (**6**) can be transformed into the following alternative forms [9]

$$\mathbf{M} = m_{00}\,\hat{\mathbf{M}}_P\,\mathbf{M}_R\,\hat{\mathbf{M}}_{\Delta D}$$
$$= m_{00}\,\mathbf{M}_R\,\hat{\mathbf{M}}'_P\,\hat{\mathbf{M}}_{\Delta D} = m_{00}\,\hat{\mathbf{M}}_P\,\hat{\mathbf{M}}'_{\Delta D}\,\mathbf{M}_R,$$
$$\left(\hat{\mathbf{M}}'_P \equiv \mathbf{M}_R^T\,\hat{\mathbf{M}}_P\,\mathbf{M}_R,\ \hat{\mathbf{M}}'_{\Delta D} \equiv \mathbf{M}_R\,\hat{\mathbf{M}}_{\Delta D}\,\mathbf{M}_R^T\right) \tag{7}$$

Note that, as indicated above, the normalized diattenuators $\hat{\mathbf{M}}_D$ and $\hat{\mathbf{M}}_P$ are completely determined by vectors $\mathbf{D}$ and $\mathbf{P}$ respectively, whereas the calculation of $\hat{\mathbf{M}}_{\Delta P}$, $\hat{\mathbf{M}}_{\Delta D}$ and $\mathbf{M}_R$ requires to distinguish between nonsingular and singular Mueller matrices. The general procedure for the calculation of $\hat{\mathbf{M}}_{\Delta P}$, $\mathbf{M}_R$ and $\hat{\mathbf{M}}_D$, including the case of $\mathbf{M}$ being singular, can be found in Ref. [2]. The calculation of $\hat{\mathbf{M}}_{\Delta D}$, $\mathbf{M}_R$ and $\hat{\mathbf{M}}_P$ is achieved by applying the same procedure to $\mathbf{M}^T$.

### B Reduced forms of a Mueller matrix

The forward Lu-Chipman decomposition (**3**) $\mathbf{M} = m_{00}\,\hat{\mathbf{M}}_{\Delta P}\,\mathbf{M}_R\,\hat{\mathbf{M}}_D$ can be modified through the diagonalization $\mathbf{m}_{\Delta P} = \mathbf{m}_{Rf}\,\mathbf{m}_{L\Delta P}\,\mathbf{m}_{Rf}^T$ of the symmetric matrix $\mathbf{m}_{\Delta P}$:

$$\mathbf{M} = m_{00}\left(\mathbf{M}_{Rf}\,\hat{\mathbf{M}}_{L\Delta P}\,\mathbf{M}_{Rf}^T\right)\mathbf{M}_R\,\hat{\mathbf{M}}_D = m_{00}\,\mathbf{M}_{Rf}\,\hat{\mathbf{M}}_{L\Delta P}\,\hat{\mathbf{M}}_{JD},$$

$$\hat{\mathbf{M}}_{L\Delta P} \equiv \begin{pmatrix} 1 & \mathbf{0}^T \\ \mathbf{P}_{L\Delta P} & \mathbf{m}_{L\Delta P} \end{pmatrix},\quad \mathbf{M}_{Rf} \equiv \begin{pmatrix} 1 & \mathbf{0}^T \\ \mathbf{0} & \mathbf{m}_{Rf} \end{pmatrix},$$

$$\mathbf{m}_{L\Delta P} \equiv \mathrm{diag}(l_{\Delta P1}, l_{\Delta P2}, \varepsilon\,l_{\Delta P3}),$$

$$\varepsilon \equiv \frac{\det \mathbf{M}}{|\det \mathbf{M}|},\ \ 0 \le l_{\Delta P3} \le l_{\Delta P2} \le l_{\Delta P1},$$

$$\mathbf{P}_{L\Delta P} = \frac{1}{1 - D^2}\mathbf{m}_{Rf}^T\left(\mathbf{P} - \mathbf{m}\mathbf{D}\right),\quad \hat{\mathbf{M}}_{JD} \equiv \mathbf{M}_{Rf}^T\,\mathbf{M}_R\,\hat{\mathbf{M}}_D, \tag{8}$$

where the only nonpure component is the depolarizer $\hat{\mathbf{M}}_{L\Delta P}$, called the *reduced form* of $\mathbf{M}$ [10], whose only nonzero elements are located on its diagonal and on its first column.

Analogously, the reverse decomposition can be expressed as follows

$$\mathbf{M} = m_{00}\, \hat{\mathbf{M}}_P\, \mathbf{M}_R \left( \mathbf{M}_{Rr}\, \hat{\mathbf{M}}_{L\Delta D}\, \mathbf{M}_{Rr}^T \right)$$

$$= m_{00}\, \hat{\mathbf{M}}_{JP}\, \hat{\mathbf{M}}_{L\Delta D}\, \mathbf{M}_{Rr}^T,$$

$$\hat{\mathbf{M}}_{L\Delta D} \equiv \begin{pmatrix} 1 & D_{L\Delta D}^T \\ \mathbf{0} & \mathbf{m}_{L\Delta D} \end{pmatrix}, \quad \mathbf{M}_{Rr} \equiv \begin{pmatrix} 1 & \mathbf{0}^T \\ \mathbf{0} & \mathbf{m}_{Rr} \end{pmatrix},$$

$$\mathbf{m}_{L\Delta D} \equiv \mathrm{diag}\left( l_{\Delta D1}, l_{\Delta D2}, \varepsilon\, l_{\Delta D3} \right),$$

$$\varepsilon \equiv \frac{\det \mathbf{M}}{|\det \mathbf{M}|}, \quad 0 \le l_{\Delta D3} \le l_{\Delta D2} \le l_{\Delta D1},$$

$$\mathbf{D}_{L\Delta D} = \frac{1}{1-P^2}\, \mathbf{m}_{Rr}\left( \mathbf{D} - \mathbf{m}^T \mathbf{P} \right),$$

$$\hat{\mathbf{M}}_{JP} \equiv \hat{\mathbf{M}}_P\, \mathbf{M}_R\, \mathbf{M}_{Rr}, \tag{9}$$

where the only nonpure component is the depolarizer $\hat{\mathbf{M}}_{L\Delta D}$, called the *reverse reduced form* of $\mathbf{M}$ [10], whose only nonzero elements are located on its diagonal and on its first row.

**C Symmetric decomposition**

Since the decompositions considered in the previous sections do not feature symmetric ordering of the components, they adopt two, forward and reverse, alternative forms. In fact, the forward ($\hat{\mathbf{M}}_{\Delta P}$ and $\hat{\mathbf{M}}_{L\Delta P}$) and reverse ($\hat{\mathbf{M}}_{\Delta D}$ and $\hat{\mathbf{M}}_{L\Delta D}$) depolarizers have asymmetric structures. Thus, it is interesting to explore the symmetric decomposition, formulated in a unique form where the depolarizer is defined without the necessity of any *a priori* choice of the order.

Any depolarizing Mueller matrix $\mathbf{M}$ can be expressed through the *symmetric decomposition* [11]

$$\mathbf{M} = \mathbf{M}_{J2}\, \mathbf{M}_\Delta\, \mathbf{M}_{J1} \tag{10}$$

where $\mathbf{M}_{J1}$ and $\mathbf{M}_{J2}$ are pure Mueller matrices. Depending on whether the N-matrix $\mathbf{N} \equiv \mathbf{G}\, \mathbf{M}^T\, \mathbf{G}\, \mathbf{M}$ [$\mathbf{G} \equiv \mathrm{diag}(1,-1,-1,-1)$] associated with $\mathbf{M}$ is diagonalizable or not, $\mathbf{M}_\Delta$ has the form of one of the two alternative canonical depolarizing Mueller matrices $\mathbf{M}_{\Delta d}$ and $\mathbf{M}_{\Delta nd}$ [12-17]

$$\mathbf{M}_{\Delta d} \equiv \mathrm{diag}(d_0, d_1, d_2, \varepsilon d_3);$$

$$d_0 \ge d_i \ge 0 \quad (i=1,2,3), \quad \varepsilon \equiv (\det \mathbf{M})/|\det \mathbf{M}|,$$

$$d_0 + d_1 \ge d_2 + \varepsilon d_3, \quad d_0 - d_1 \ge d_2 - \varepsilon d_3, \tag{11}$$

$$\mathbf{M}_{\Delta nd} \equiv \begin{pmatrix} 2a_0 & -a_0 & 0 & 0 \\ a_0 & 0 & 0 & 0 \\ 0 & 0 & a_2 & 0 \\ 0 & 0 & 0 & a_2 \end{pmatrix}; \quad (0 \le a_2 \le a_0) \tag{12}$$

The parameters $(d_0, d_1, d_2, d_3)$ and $(a_0, a_0, a_2, a_2)$ are the square roots of the (nonnegative) eigenvalues $(\rho_0, \rho_1, \rho_2, \rho_3)$ of $\mathbf{N}$.

Let us recall that $\mathbf{M}$ is called of type-I when its associated N-Matrix $\mathbf{N}$ is diagonalizable; otherwise $\mathbf{M}$ is of type-II. An alternative (but equivalent) interesting criterion to distinguish type-I and type-II Mueller matrices is obtained by considering the following pair of eigenvalue-eigenvector equations

$$\mathbf{N}\hat{\mathbf{s}}_{Pp} = c_0^2\, \hat{\mathbf{s}}_{Pp}, \quad \mathbf{N}^T \hat{\mathbf{s}}_{Dp} = c_0^2\, \hat{\mathbf{s}}_{Dp};$$

$$\text{with} \quad \hat{\mathbf{s}}_{Pp} \equiv \begin{pmatrix} 1 \\ \mathbf{P}_p \end{pmatrix}, \; \hat{\mathbf{s}}_{Dp} \equiv \begin{pmatrix} 1 \\ \mathbf{D}_p \end{pmatrix}, \tag{13}$$

where $\hat{\mathbf{s}}_{Pp}$ and $\hat{\mathbf{s}}_{Dp}$ are normalized Stokes vectors; it can be shown that $\mathbf{M}$ is type-II if and only if $\hat{\mathbf{s}}_{Pp}$ and $\hat{\mathbf{s}}_{Dp}$ correspond to totally polarized states (i.e., $|\mathbf{P}_p| = |\mathbf{D}_p| = 1$).

The procedure for the calculation of the components of the symmetric decomposition, including the case of $\mathbf{M}$ being singular, can be found in Refs. [11,18].

## 3. CHARACTERISTIC ELLIPSOIDS OF A MUELLER MATRIX

The combination of the above serial decompositions and the geometric representation of Mueller matrices through representative ellipsoids provides a privileged framework for the analysis and interpretation of the different types of singular Mueller matrices. This section is devoted to the identification and parameterization of the characteristic ellipsoids as a necessary step prior to the case study dealt with in the following section. Normalized Stokes vectors are commonly parameterized as

$$\hat{\mathbf{s}} \equiv \frac{\mathbf{s}}{s_0} \equiv \begin{pmatrix} 1 \\ \hat{s}_1 \\ \hat{s}_2 \\ \hat{s}_3 \end{pmatrix} \equiv \begin{pmatrix} 1 \\ \mathcal{P}\mathbf{u} \end{pmatrix} \equiv \begin{pmatrix} 1 \\ \mathbf{p} \end{pmatrix};$$

$$\mathbf{u} \equiv \begin{pmatrix} u_1 \\ u_2 \\ u_3 \end{pmatrix} \equiv \begin{pmatrix} \cos 2\chi \cos 2\varphi \\ \cos 2\chi \sin 2\varphi \\ \sin 2\chi \end{pmatrix}, \tag{14}$$

where $\varphi$ is the azimuth $(0 \le \varphi < \pi)$ and $\chi$ is the ellipticity angle $(-\pi/4 \le \chi \le \pi/4)$ of the *characteristic polarization state* (i.e., of the average polarization ellipse) and $\mathcal{P}$ is the degree of polarization. The Poincaré sphere is the solid sphere constituted by the points defined by the entire set of *polarization vectors* (or *Pauli vectors*) $\mathbf{p} \equiv (\hat{s}_1, \hat{s}_2, \hat{s}_3) = (\mathcal{P}u_1, \mathcal{P}u_2, \mathcal{P}u_3)$.

Normalized totally polarized states are denoted as

$$\hat{\mathbf{s}}_p \equiv \frac{\mathbf{s}_p}{s_0} \equiv (1, \hat{s}_{p1}, \hat{s}_{p2}, \hat{s}_{p3})^T \equiv \begin{pmatrix} 1 \\ \mathbf{u} \end{pmatrix} \equiv \begin{pmatrix} 1 \\ \mathbf{p}_p \end{pmatrix}. \tag{15}$$

The *P*-image of **M** is defined as the surface constituted by points

$$\mathbf{p}'_p \equiv \frac{1}{(\mathbf{M}\hat{\mathbf{s}}_p)_0} \left[ (\mathbf{M}\hat{\mathbf{s}}_p)_1, (\mathbf{M}\hat{\mathbf{s}}_p)_2, (\mathbf{M}\hat{\mathbf{s}}_p)_3 \right]^T. \tag{16}$$

Since $\mathbf{p}'_p$ is independent of $m_{00}$, the normalized Mueller matrix $\hat{\mathbf{M}} \equiv \mathbf{M}/m_{00}$ is an appropriate representative of $\mathbf{M}$ when dealing with the *P*-image.

It has been demonstrated that the *P*-image of a given Mueller matrix $\mathbf{M}$ is an ellipsoid [19,20], hereafter called the *forward ellipsoid* $E_{\Delta P}$. Furthermore, since the transposed Mueller matrix $\mathbf{M}^T$ is also a Mueller matrix [3,7,20-22], it has also an associated ellipsoid $E_{\Delta D}$, hereafter called the *reverse ellipsoid* of $\mathbf{M}$. Note that, since a physical Mueller matrix does not *overpolarize* (i.e., satisfies the Stokes criterion [12]), the *representative ellipsoids* $E_{\Delta P}$ and $E_{\Delta D}$ do not protrude outside the unit sphere.

Let us now consider the forward decomposition

$$\mathbf{M} = m_{00}\, \mathbf{M}_{Rf}\, \hat{\mathbf{M}}_{L\Delta P}\, \hat{\mathbf{M}}_{JD}. \tag{17}$$

expressed in terms of the forward reduced form $\hat{\mathbf{M}}_{L\Delta P}$ of $\mathbf{M}$ and observe that, leaving aside the case of $\hat{\mathbf{M}}_D$ being singular (hence, $\hat{\mathbf{M}}_{JD}$ being singular), the *entrance pure Mueller matrix* $\hat{\mathbf{M}}_{JD} \equiv \mathbf{M}_R^T\, \mathbf{M}_R\, \hat{\mathbf{M}}_D$ maps the Poincaré sphere onto itself. Furthermore, $\hat{\mathbf{M}}_{L\Delta P}$ can be written as [19]

$$\hat{\mathbf{M}}_{L\Delta P} \equiv \begin{pmatrix} 1 & \mathbf{0}^T \\ \mathbf{P}_{L\Delta P} & \mathbf{I}_3 \end{pmatrix} \begin{pmatrix} 1 & \mathbf{0}^T \\ \mathbf{0} & \mathbf{m}_{\Delta P} \end{pmatrix}, \tag{18}$$

so that, for any input totally polarized Stokes vector $\hat{\mathbf{s}}_p$, the output Stokes vector $\mathbf{s}'$ provided by $\hat{\mathbf{M}}_{L\Delta P}$ is given by

$$\hat{\mathbf{M}}_{L\Delta P}\, \hat{\mathbf{s}}_p = \begin{pmatrix} 1 \\ \mathbf{P}_{L\Delta P} + \mathbf{m}_{\Delta P}\mathbf{u} \end{pmatrix} = \begin{pmatrix} 0 \\ \mathbf{P}_{L\Delta P} \end{pmatrix} + \begin{pmatrix} 1 \\ \mathbf{m}_{\Delta P}\mathbf{u} \end{pmatrix}. \tag{19}$$

The diagonal right matrix factor in Eq. (**18**) (representing an *intrinsic depolarizer*) maps the Poincaré sphere onto a centered ellipsoid whose semiaxes are given by the absolute values of the diagonal elements of the diagonal matrix $\mathbf{m}_{L\Delta P}$, while the directions of the principal axes are given by the reference axes $S_1 S_2 S_3$ themselves. The left matrix factor induces geometrically a global shift $\mathbf{P}_{L\Delta P}$ of the ellipsoid.

Provided the diattenuation $D$ of $\mathbf{M}$ is less than one, the *P*-image of the normalized nonsingular pure Mueller matrix $\hat{\mathbf{M}}_{JD}$ is the unit sphere itself and, consequently, the *P*-image of $\hat{\mathbf{M}}_{L\Delta P}\, \hat{\mathbf{M}}_{JD}$ coincides with that of $\hat{\mathbf{M}}_{L\Delta P}$, so that the unit sphere is transformed into an ellipsoid $E_{L\Delta P}$ (located inside the unit sphere). Its principal axes are aligned with the Poincaré's reference frame; its semiaxes are given by $(l_{\Delta P1}, l_{\Delta P2}, l_{\Delta P3})$ and its displacement from the origin of the sphere is given by vector $\mathbf{P}_{L\Delta P}$ ($\mathbf{P}_{L\Delta P}$ runs from the center of the Poincaré sphere to the center of the ellipsoid $E_{L\Delta P}$). The complete mapping effect of $\mathbf{M}$ is finally obtained through the additional action of the left factor matrix $\mathbf{M}_{Rf}$ in Eq. (**8**) which consists in applying a rigid rotation of both $\mathbf{P}_{L\Delta P}$ and $E_{L\Delta P}$ by the angle

$$\Delta_{\Delta P} = \arccos\left[(\mathrm{tr}\, \mathbf{M}_{Rf})/2 - 1\right] \tag{20}$$

about the axis determined by the normalized retardance vector:

$$\hat{\mathbf{R}}_f \equiv \left(\cos 2\chi_{\Delta P} \cos 2\varphi_{\Delta P},\, \cos 2\chi_{\Delta P} \sin 2\varphi_{\Delta P},\, \sin 2\chi_{\Delta P}\right)^T. \tag{21}$$

In summary, the *P*-image of $\mathbf{M}$ results in a *forward ellipsoid* $E_{\Delta P}$ with semiaxes $(l_{\Delta P1}, l_{\Delta P2}, l_{\Delta P3})$, which is tilted with respect to the Poincaré's reference frame $S_1 S_2 S_3$ by the rotation effect produced by $\mathbf{M}_{Rf}$ and whose displacement from the central position is given by vector $\mathbf{P}_{\Delta P}$ (note that $\mathbf{P}_{\Delta P} = \mathbf{m}_{Rf}\, \mathbf{P}_{L\Delta P}$).

The complete geometric information involved in the forward ellipsoid $E_{\Delta P}$ consists of nine independent parameters, namely: the three semiaxes $(l_{\Delta P1}, l_{\Delta P2}, l_{\Delta P3})$ of $E_{\Delta P}$; the three orientation angles $(\varphi_{\Delta P}, \chi_{\Delta P}, \Delta_{\Delta P})$ of the axes of $E_{\Delta P}$ with respect to the Poincaré's reference frame $S_1 S_2 S_3$, and the three coordinates of the geometric center of $E_{\Delta P}$ (with respect to $S_1 S_2 S_3$), given by the components $(P_{\Delta P1}, P_{\Delta P2}, P_{\Delta P3})$ of the polarizance vector $\mathbf{P}_{\Delta P}$.

These nine geometric parameters determining the shape and the location of $E_{\Delta P}$, do not provide sufficient information to recover all the fifteen elements of $\hat{\mathbf{M}}$. Nevertheless, complementary information can be obtained by additionally considering the *P*-image of $\hat{\mathbf{M}}^T$ [20]. Since $\hat{\mathbf{M}}^T$ can be expressed as $\hat{\mathbf{M}}^T = \hat{\mathbf{M}}_{\Delta D}^T \mathbf{M}_R^T \hat{\mathbf{M}}_P = \mathbf{M}_{Rr} \hat{\mathbf{M}}_{L\Delta D}^T \hat{\mathbf{M}}_{JP}^T$ (with $\hat{\mathbf{M}}_{JP}^T \equiv \hat{\mathbf{M}}_{Rr}^T \mathbf{M}_R^T \hat{\mathbf{M}}_P$) and noticing that the *P*-image of the pure Mueller matrix $\hat{\mathbf{M}}_{JP}^T$ (assumed nonsingular) is the unit sphere itself, the *P*-image of $\hat{\mathbf{M}}^T$ can be obtained through the consecutive effects of $\hat{\mathbf{M}}_{L\Delta D}^T$ followed by the retarder matrix $\mathbf{M}_{Rr}$, whose mapping results in defining a *reverse ellipsoid* $E_{\Delta D}$, dual to $E_{\Delta P}$ and likewise located inside the unit sphere. The semiaxes of $E_{\Delta D}$ are given by $(l_{\Delta D1}, l_{\Delta D2}, l_{\Delta D3})$, its principal axes are tilted with respect to the Poincaré's reference frame $S_1 S_2 S_3$ by the effect of the rotation matrix $\mathbf{m}_{Rr}$, and its displacement from the central position is determined by vector $\mathbf{D}_{\Delta D}$.

Consequently, the pair of *representative ellipsoids* $E_{\Delta P}$ and $E_{\Delta D}$ provides a geometric representation of a given normalized depolarizing Mueller matrix $\hat{\mathbf{M}}$ that enables to recover all the information involved in $\hat{\mathbf{M}}$. Obviously, there exist certain constraining relations between the nine geometric parameters of $E_{\Delta P}$ and those of $E_{\Delta D}$ that reduce the number of independent geometric quantities to fifteen [20].

Leaving aside the fact that the *canonical ellipsoid* $E_\Delta$ associated with the central canonical depolarizer of the symmetric decomposition of $\hat{\mathbf{M}}$ [20] can always be obtained from $\hat{\mathbf{M}}$, and hence from $E_{\Delta P}$ and $E_{\Delta D}$, $E_\Delta$ further provides a useful geometric viewpoint on certain properties of $\hat{\mathbf{M}}$. Consequently, for the sake of practical and meaningful geometric characterization of $\hat{\mathbf{M}}$, it is worth considering jointly the three associated *characteristic ellipsoids* $E_{\Delta P}$, $E_{\Delta D}$ and $E_\Delta$. The specific canonical ellipsoids associated with the type-I and type-II central canonical depolarizers $\hat{\mathbf{M}}_{\Delta d}$ and $\hat{\mathbf{M}}_{\Delta nd}$, whose properties have been studied in detail in Ref. [20], are respectively denoted $E_{\Delta d}$ and $E_{\Delta nd}$. The semiaxes of $E_{\Delta d}$ (which is a centered ellipsoid) are $(\hat{d}_1, \hat{d}_2, \hat{d}_3)$, with $\hat{d}_i \equiv d_i/d_0$ ($i = 1, 2, 3$). The semiaxes of the type-II canonical ellipsoid $E_{\Delta nd}$ (which is an ellipsoid of revolution) are $(1/3, a_2/\sqrt{3}\, a_0, a_2/\sqrt{3}\, a_0)$; the coordinates of its centre are $(2/3, 0, 0)$, so that $E_{\Delta nd}$ has the peculiarity that it touches the unit sphere at the single point $(1, 0, 0)$ [20].

Since $\mathbf{M}_{J2}$ is a pure Mueller matrix, it does not depolarize totally polarized input states, and consequently, points of $E_\Delta$ located on the surface of the unit sphere are transformed into points of $E_{\Delta P}$ lying also on the surface. Furthermore, when $P < 1$ (i.e., $\det \mathbf{M}_{J2} > 0$) points of $E_\Delta$ that are not located on the surface of the unit sphere are transformed into points of $E_{\Delta P}$ that are inside the unit sphere and not on its surface. The same arguments are valid for $E_{\Delta D}$: when $D < 1$ (i.e., $\det \mathbf{M}_{J1} > 0$) the only points of $E_{\Delta D}$ touching the unit sphere are the contact image points of $E_\Delta$ through $\mathbf{M}_{J1}^T$.

## 4. SINGULAR MUELLER MATRICES

From Eqs. (**8**) and (**9**) the determinant of $\mathbf{M}$ can be expressed in the following forms

$$\det \mathbf{M} = m_{00}^4\, \det \hat{\mathbf{M}}_{L\Delta P}\, \det \hat{\mathbf{M}}_D = \varepsilon\, m_{00}^4\, l_{\Delta P1}\, l_{\Delta P2}\, l_{\Delta P3}\, (1 - D^2),$$

$$\det \mathbf{M} = m_{00}^4\, \det \hat{\mathbf{M}}_P\, \det \hat{\mathbf{M}}_{L\Delta D} = \varepsilon\, m_{00}^4\, l_{\Delta D1}\, l_{\Delta D2}\, l_{\Delta D3}\, (1 - P^2), \tag{22}$$

and consequently, leaving aside the trivial case of the zero Mueller matrix (i.e., $m_{00} = 0$), $\mathbf{M}$ satisfies $\det \mathbf{M} = 0$ if and only if at least one of the following conditions holds: $P = 1$ (i.e., $\hat{\mathbf{M}}$ corresponds to a *depolarizing polarizer* [2]); $D = 1$ ($\hat{\mathbf{M}}$ corresponds to a *depolarizing analyzer* [2]), or $l_{\Delta P3} = l_{\Delta D3} = 0$ (in which case the medium will be called a *singular depolarizer*). Recall that, when $P = D = 1$ the medium is necessarily a nondepolarizing polarizer (also called *polarizer-analyzer*, or simply *polarizer*).

In the light of the serial decompositions considered, the analysis and classification of singular Mueller matrices can be performed through the study of separate cases according to the values of the polarizance $P$ and the diattenuation $D$ of $\mathbf{M}$: *a)* $P=1$ and $D<1$; *b)* $D=1$ and $P<1$; *c)* $P=D=1$, and *d)* $D<1$ and $P<1$. The geometric features of depolarizing polarizers (*a*), depolarizing analyzers (*b*), and polarizers (*c*) are summarized in Table 1, while those of singular depolarizers are outlined in tables II and III.

As indicated in the introduction, when the calculated determinant of a measured Mueller matrix $\mathbf{M}_{\exp}$ has a sufficiently small value (that is, it is of the same order as the experimental error), then the smaller eigenvalue $\rho_3$ of the N-matrix $\mathbf{N} \equiv \mathbf{GM}_{\exp}^T\mathbf{GM}_{\exp}$ can be replaced by zero, and the filtered Mueller matrix $\mathbf{M}_f$ can be constructed from the symmetric composition $\mathbf{M}_f = \mathbf{M}_{J2}\mathbf{M}_{\Delta f}\mathbf{M}_{J1}$, where the pure components $\mathbf{M}_{J1}$ and $\mathbf{M}_{J2}$ are those of the symmetric decomposition $\mathbf{M}_{\exp} = \mathbf{M}_{J2}\mathbf{M}_{\Delta}\mathbf{M}_{J1}$ and $\mathbf{M}_{\Delta f}$ is built from $\mathbf{M}_{\Delta}$ but replacing $d_3 = \sqrt{\rho_3}$ by zero (if $\mathbf{M}_{\exp}$ is type-I) or replacing $a_2 = \sqrt{\rho_3}$ by zero (if $\mathbf{M}_{\exp}$ is type-II), so that the condition $\det \mathbf{M}_f = 0$ is satisfied exactly.

### A. Depolarizing polarizer

When $P=1$ and $D<1$, the Mueller matrix is necessarily singular and of type-I [11] and corresponds to a depolarizing polarizer, which has the characteristic form

$$\mathbf{M}_{\hat{P}D} = m_{00}\begin{pmatrix} 1 & \mathbf{D}^T \\ \hat{\mathbf{P}} & \hat{\mathbf{P}}\otimes\mathbf{D}^T \end{pmatrix};\; P\equiv|\hat{\mathbf{P}}|=1,\; D\equiv|\mathbf{D}|<1, \quad (23)$$

so that the forward ellipsoid $E_{\Delta P}$ degenerates into a single point located on the surface of the unit sphere at the end-point of the unit polarizance vector $\hat{\mathbf{P}}$ (regardless of the value of $D$) and the reverse ellipsoid $E_{\Delta D}$ degenerates into a single point located inside the unit sphere at the end-point of the diattenuation vector **D**.

Note that a depolarizing polarizer $\mathbf{M}_{\hat{P}D}$ depends on six independent parameters specified by $m_{00}$, $\hat{\mathbf{P}}$ (with $P\equiv|\hat{\mathbf{P}}|=1$) and **D**. Furthermore, the normalized input and output diattenuators $\hat{\mathbf{M}}_{D1}$ and $\hat{\mathbf{M}}_{D2}$ of the symmetric decomposition coincide respectively with $\hat{\mathbf{M}}_{D}$ and $\hat{\mathbf{M}}_{\hat{P}}$ [see Eqs. (**4**) and (**5**)] and therefore, $P = D_2 = 1$ and $D = D_1 < 1$.

The submatrices $\mathbf{m}_{L\Delta P}$, $\mathbf{m}_{L\Delta D}$ and $\mathbf{m}_{\Delta d}$, corresponding respectively to the forward and reverse reduced forms and to the canonical depolarizer of $\mathbf{M}_{\hat{P}D}$, equal the zero matrix,

$$\mathbf{m}_{L\Delta P}\left(\mathbf{M}_{\hat{P}D}\right) = \mathbf{m}_{L\Delta D}\left(\mathbf{M}_{\hat{P}D}\right) = \mathbf{m}_{\Delta d}\left(\mathbf{M}_{\hat{P}D}\right) = \mathrm{diag}(0,0,0) \quad (24)$$

Therefore, the type-I canonical ellipsoid $E_{\Delta d}$ of $\mathbf{M}_{\hat{P}D}$ degenerates into a single point located at the origin.

Depolarizing polarizers are characterized by the fact that the output state of polarization is fixed and totally polarized, so that any serial combination of an arbitrary depolarizing medium followed by a polarizer produces a depolarizing polarizer. As an example, let us consider the serial combination $\mathbf{M}_{\hat{D}}\mathbf{M}_{\Delta I}$ where $\mathbf{M}_{\hat{D}}$ corresponds to a normal polarizer and $\mathbf{M}_{\Delta I} \equiv \mathrm{diag}(1,a,a,a)$ corresponds to an *intrinsic isotropic depolarizer*, so that the resulting depolarizing polarizer has the particular form

$$\mathbf{M}_{\hat{D}}\mathbf{M}_{\Delta I} = m_{00}\begin{pmatrix} 1 & \hat{\mathbf{D}}^T \\ \hat{\mathbf{D}} & \hat{\mathbf{D}}\otimes\hat{\mathbf{D}}^T \end{pmatrix}\begin{pmatrix} 1 & \mathbf{0}^T \\ \mathbf{0} & a\mathbf{I}_3 \end{pmatrix} = m_{00}\begin{pmatrix} 1 & a\hat{\mathbf{D}}^T \\ \hat{\mathbf{D}} & a\hat{\mathbf{D}}\otimes\hat{\mathbf{D}}^T \end{pmatrix} \quad (25)$$

Furthermore, if we consider an arbitrary input polarization state with Stokes vector

$$\mathbf{s} \equiv I\begin{pmatrix} 1 \\ P\hat{\mathbf{u}} \end{pmatrix}, \quad (26)$$

then **s** is transformed as follows

$$\left(\mathbf{M}_{\hat{D}}\mathbf{M}_{\Delta I}\right)I\begin{pmatrix} 1 \\ P\hat{\mathbf{u}} \end{pmatrix} = m_{00}I\left(1+aP\hat{\mathbf{D}}^T\hat{\mathbf{u}}\right)\begin{pmatrix} 1 \\ \hat{\mathbf{D}} \end{pmatrix}. \quad (27)$$

Observe that, even when $a=0$, the characteristic features of $\mathbf{M}_{\hat{D}}\mathbf{M}_{\Delta I}$ being a depolarizing polarizer are maintained. The above example reflects the general property of depolarizing polarizers that, despite the fact that the degree of polarization of the output state is equal to one, the depolarization index $P_\Delta$ is less that one because of the presence of the depolarizer preceding the polarizer. The property $P_\Delta < 1$ of a depolarizing polarizer $\mathbf{M}_{\hat{P}D}$ is intrinsically related to the fact that its reverse matrix of $\mathbf{M}^r_{\hat{P}D}$ depolarizes (to some extent) totally polarized input states; in fact $\mathbf{M}^r_{\hat{P}D}$ is a depolarizing analyzer that, as shown in the next Section, produces a partially polarized output state regardless of the state of polarization of the input beam.

### B. Depolarizing analyzer

When $P<1$ and $D=1$, the Mueller matrix is necessarily singular and of type-I [11] and corresponds to a depolarizing analyzer with the characteristic form

$$\mathbf{M}_{P\hat{D}} = m_{00}\begin{pmatrix} 1 & \hat{\mathbf{D}}^T \\ \mathbf{P} & \mathbf{P}\otimes\hat{\mathbf{D}}^T \end{pmatrix};\; P\equiv|\mathbf{P}|<1,\; D\equiv|\hat{\mathbf{D}}|=1, \quad (28)$$

and the forward ellipsoid $E_{\Delta P}$ degenerates into a single point located inside the unit sphere at the end-point of the polarizance vector **P**, whereas the reverse ellipsoid $E_{\Delta D}$ degenerates into a single point located at the end-point of the unit diattenuation vector $\hat{\mathbf{D}}$, located on the surface of the unit sphere (regardless of the value of $P$).

The Mueller matrix $\mathbf{M}_{P\hat{D}}$ of a depolarizing analyzer depends on six independent parameters determined by $m_{00}$, $\hat{\mathbf{D}}$ (with $D\equiv|\hat{\mathbf{D}}|=1$) and **P**. The normalized input and output diattenuators $\hat{\mathbf{M}}_{D1}$ and $\hat{\mathbf{M}}_{D2}$ of the symmetric decomposition of $\mathbf{M}_{P\hat{D}}$ coincide respectively with $\hat{\mathbf{M}}_{\hat{D}}$ and $\hat{\mathbf{M}}_P$, so that $D = D_1 = 1$ and $P = D_2 < 1$.

Thus, as in the case of depolarizing polarizers, the diagonal submatrices $\mathbf{m}_{L\Delta P}$, $\mathbf{m}_{L\Delta D}$ and $\mathbf{m}_{\Delta d}$ coincide with the zero matrix,

$$\mathbf{m}_{L\Delta P}\left(\mathbf{M}_{P\hat{D}}\right) = \mathbf{m}_{L\Delta D}\left(\mathbf{M}_{P\hat{D}}\right) = \mathbf{m}_{\Delta d}\left(\mathbf{M}_{P\hat{D}}\right) = \mathrm{diag}(0,0,0) \quad (29)$$

so that the type-I canonical ellipsoid $E_{\Delta d}$ of $\mathbf{M}_{P\hat{D}}$ degenerates into a single point located at the origin.

Depolarizing analyzers are characterized by the fact that the output state of polarization is fixed and partially polarized, so that any serial combination of a polarizer followed by an arbitrary depolarizing medium produces a depolarizing analyzer. As an example, let us consider the serial combination $\mathbf{M}_{\Delta I}\mathbf{M}_{\hat{D}}$, so that the resulting depolarizing analyzer has the particular form

$$\mathbf{M}_{\Delta I}\mathbf{M}_{\hat{D}} = \begin{pmatrix} 1 & \mathbf{0}^T \\ \mathbf{0} & a\mathbf{I}_3 \end{pmatrix}m_{00}\begin{pmatrix} 1 & \hat{\mathbf{D}}^T \\ \hat{\mathbf{D}} & \hat{\mathbf{D}}\otimes\hat{\mathbf{D}}^T \end{pmatrix} = m_{00}\begin{pmatrix} 1 & \hat{\mathbf{D}}^T \\ a\hat{\mathbf{D}} & a\hat{\mathbf{D}}\otimes\hat{\mathbf{D}}^T \end{pmatrix} \quad (30)$$

and transforms an arbitrary input polarization state **s** in the following manner

$$\left(\mathbf{M}_{\Delta I}\mathbf{M}_{\hat{D}}\right)I\begin{pmatrix} 1 \\ P\hat{\mathbf{u}} \end{pmatrix} = m_{00}I\left(1+P\hat{\mathbf{D}}^T\hat{\mathbf{u}}\right)\begin{pmatrix} 1 \\ a\hat{\mathbf{D}} \end{pmatrix}. \quad (31)$$

Thus, even when $a=0$, the characteristic features of $\mathbf{M}_{\Delta I}\mathbf{M}_{\hat{D}}$ being a depolarizing analyzer are maintained. The above example reflects the general property of depolarizing analyzers that, despite that they can be considered as containing a polarizer as serial component (followed by a depolarizing component), their depolarization index $P_\Delta$ is always less than one.

### C. Polarizer

When $P = D = 1$ the Mueller matrix is pure (and hence, is of type-I) and corresponds to a polarizer, which, in general, is *non-normal* [22,23], or *inhomogeneous* [24], in the sense that the equality $\mathbf{M}^T\mathbf{M} = \mathbf{M}\mathbf{M}^T$ does not necessarily hold. The Mueller matrix of any polarizer has the form

$$\mathbf{M}_{\hat{P}\hat{D}} = m_{00} \begin{pmatrix} 1 & \hat{\mathbf{D}}^T \\ \hat{\mathbf{P}} & \hat{\mathbf{P}} \otimes \hat{\mathbf{D}}^T \end{pmatrix} \quad (P = D = 1). \tag{32}$$

The forward and reverse depolarizers $\hat{\mathbf{M}}_{\Delta P}$ and $\hat{\mathbf{M}}_{\Delta D}$, as well as the forward and reverse reduced forms $\hat{\mathbf{M}}_{L\Delta P}$ and $\hat{\mathbf{M}}_{L\Delta D}$ of any polarizer are given by the identity matrix, so that

$$\mathbf{m}_{L\Delta P}(\mathbf{M}_{\hat{P}\hat{D}}) = \mathbf{m}_{L\Delta D}(\mathbf{M}_{\hat{P}\hat{D}}) = \mathbf{I}_3. \tag{33}$$

Due to the peculiar properties of $\hat{\mathbf{M}}_{\hat{P}\hat{D}}$ (pure and singular), its symmetric decomposition can be taken as $\mathbf{M}_{\hat{P}\hat{D}} = \mathbf{M}_{\hat{P}} \mathbf{M}_{\Delta 0} \mathbf{M}_{\hat{D}}$, where $\mathbf{M}_{\Delta 0} \equiv \text{diag}(1,0,0,0)$ represents an ideal depolarizer, or, except for the case that $\mathbf{P} = -\mathbf{D}$ (degenerate polarizer), the symmetric decomposition $\mathbf{M}_{\hat{P}\hat{D}} = \mathbf{M}_{\hat{P}} \mathbf{I}_4 \mathbf{M}_{\hat{D}}$ ($\mathbf{I}_4$ being the 4x4 identity matrix) is also physically admissible, and consequently the canonical ellipsoid of $\hat{\mathbf{M}}_{\hat{P}\hat{D}}$ is undetermined. In fact, when the rightmost and leftmost components of a serial combination of Mueller matrices are both polarizers, the composed medium necessarily behaves as a polarizer, and even the sandwich

$$\mathbf{M}_{\hat{P}} \mathbf{M}_{\Delta I} \mathbf{M}_{\hat{D}} = \frac{1}{2}\begin{pmatrix} 1 & \hat{\mathbf{P}}^T \\ \hat{\mathbf{P}} & \hat{\mathbf{P}} \otimes \hat{\mathbf{P}}^T \end{pmatrix} \begin{pmatrix} 1 & \mathbf{0}^T \\ \mathbf{0} & a\mathbf{I}_3 \end{pmatrix} \frac{1}{2}\begin{pmatrix} 1 & \hat{\mathbf{D}}^T \\ \hat{\mathbf{D}} & \hat{\mathbf{D}} \otimes \hat{\mathbf{D}}^T \end{pmatrix}$$

$$= \frac{1 + a\hat{\mathbf{P}}^T\hat{\mathbf{D}}}{4}\begin{pmatrix} 1 & \hat{\mathbf{D}}^T \\ \hat{\mathbf{P}} & \hat{\mathbf{P}} \otimes \hat{\mathbf{D}}^T \end{pmatrix} \quad (a < 1), \tag{34}$$

produces a polarizer, regardless of the value of *a* (with $a < 1$), and thus $P_\Delta = 1$. The representative ellipsoids $E_{\Delta P}$ and $E_{\Delta D}$ degenerate into the respective end-points of the unit vectors $\hat{\mathbf{P}}$ and $\hat{\mathbf{D}}$ located on the surface of the unit sphere.

### D. Singular depolarizer

The case study is completed by considering a *singular depolarizer*, characterized by an associated Mueller matrix $\mathbf{M}_{S\Delta}$ that, while satisfying $\det \mathbf{M}_{S\Delta} = 0$, it does not correspond either to a polarizer (i.e., $P < 1$), or to an analyzer (i.e., $D < 1$). From Eqs. (**22**), we deduce that in this case the equalities $l_{\Delta P3} = l_{\Delta D3} = 0$ must hold necessarily (while the parameters $l_{\Delta P2}$, $l_{\Delta D2}$, $l_{\Delta P1}$, $l_{\Delta D1}$, with $0 \leq l_{\Delta P2} \leq l_{\Delta P1}$ and $0 \leq l_{\Delta D2} \leq l_{\Delta D1}$, may vanish or not). Since $P < 1$ and $D < 1$, neither $\mathbf{M}_{J2}$, nor $\mathbf{M}_{J1}$ of the normal form $\mathbf{M}_{S\Delta} = \mathbf{M}_{J2} \mathbf{M}_\Delta \mathbf{M}_{J1}$ are polarizers and therefore, the number of zero semiaxes is identical for the three characteristic ellipsoids $E_{\Delta P}$, $E_{\Delta D}$ and $E_\Delta$. Thus, it is the depolarizing behavior of the medium which leads to the property $\det \mathbf{M}_{S\Delta} = 0$, rather than $\mathbf{M}_{S\Delta}$ corresponding to a polarizer or to an analyzer (depolarizing or not).

Note also that, since a depolarizing Mueller matrix can always be considered as a convex sum (or ensemble average) of pure Mueller matrices [25-28], the existence of singular depolarizers reflects the fact that certain statistical mixtures of Mueller matrices result in an overall singular Mueller matrix. A simple example of this behavior can be synthesized by means of a parallel combination of a neutral filter whose Mueller matrix is the identity matrix and a half-wave plate oriented at 0º,

$$\begin{aligned}\mathbf{M}_{S\Delta} &= \frac{1}{2}\text{diag}(1,1,1,1) + \frac{1}{2}\text{diag}(1,1,-1,-1) \\ &= \text{diag}(1,1,0,0)\end{aligned} \tag{35}$$

This matrix preserves the input intensity and exhibits both zero diattenuation and polarizance $(P = D = 0)$, but nevertheless produces output states whose circular and ±45º-linear Stokes vector components vanish (i.e., the fractions of the intensity associated with these components are totally randomized).

The symmetric decomposition (**10**) provides a particularly insightful way to analyze the properties of singular depolarizers. In fact, $0 = \det \mathbf{M}_{S\Delta} = \det \mathbf{M}_{J2} \det \mathbf{M}_\Delta \det \mathbf{M}_{J1}$ must be satisfied simultaneously with $\det \mathbf{M}_{J1} > 0$ and $\det \mathbf{M}_{J2} > 0$ and therefore, $\det \mathbf{M}_\Delta = 0$, which implies either $d_3 = 0$, if $\mathbf{M}_{S\Delta}$ is of type-I (with the common choice $d_3 \leq d_2 \leq d_1 \leq d_0$ indicated in Eq. (**11**)), or $a_2 = 0$, if $\mathbf{M}_{S\Delta}$ is of type-II.

Thus, the canonical depolarizer of a type-I singular depolarizer necessarily extinguishes the circular component (at least) of the state of polarization of the incident light, while the canonical depolarizer of a type-II singular depolarizer extinguishes both the circular and the ±45º-linear components. That is, $\mathbf{M}_{J1}$ transforms the Stokes vector **s** of the incident light into another state $\mathbf{s}_1 = \mathbf{M}_{J1}\mathbf{s}$ and then, through the action of $\mathbf{M}_\Delta$, at least the last Stokes parameter of $\mathbf{M}_\Delta\mathbf{s}_1$ vanishes when $\mathbf{M}_{S\Delta}$ is of type-I, or the last two Stokes parameters of $\mathbf{M}_\Delta\mathbf{s}_1$ vanish when $\mathbf{M}_{S\Delta}$ is of type-II; finally, $\mathbf{M}_{J2}$ acts on the state $\mathbf{M}_\Delta\mathbf{s}_1$.

In accordance with the above analysis, when considering type-I singular depolarizers, the following cases can be distinguished:

a) $d_3 = 0$, $0 < \hat{d}_2 \leq \hat{d}_1$. $E_{\Delta d}$ degenerates into an ellipse. Therefore, the forward and reverse ellipsoids degenerate into respective ellipses whose semiaxes satisfy $l_{\Delta P3} = 0$, $0 < l_{\Delta P2} \leq l_{\Delta P1}$ (for $E_{\Delta P}$) and $l_{\Delta D3} = 0$, $0 < l_{\Delta D2} \leq l_{\Delta D1}$ (for $E_{\Delta D}$).

  a1) $0 < \hat{d}_2 \leq \hat{d}_1 < 1$. The ellipses $E_{\Delta d}$, $E_{\Delta P}$ (which satisfies $0 < l_{\Delta P2} \leq l_{\Delta P1} < 1$) and $E_{\Delta D}$ (which satisfies $0 < l_{\Delta D2} \leq l_{\Delta D1} < 1$) do not touch the unit sphere at any point. The ellipse $E_{\Delta d}$ is a circle when $d_2 = d_1$ ($E_{\Delta P}$ and $E_{\Delta D}$ are not necessarily circle-shaped because $d_2 = d_1$ implies neither $l_{\Delta P2} = l_{\Delta P1}$, nor $l_{\Delta D2} = l_{\Delta D1}$).

  a2) $0 < \hat{d}_2 < \hat{d}_1 = 1$. The ellipse $E_{\Delta d}$ touches the unit sphere at only two antipodal points, while $E_{\Delta P}$ and $E_{\Delta D}$ likewise touch the sphere at respective pairs of points only (not mutually antipodal in general, due to the diattenuating effects of $\mathbf{M}_{J2}$ and $\mathbf{M}_{J1}$).

  a3) $\hat{d}_2 = \hat{d}_1 = 1$. This case is unphysical because the matrix $\text{diag}(1,1,1,0)$ does not satisfy the required inequalities (**11**) and consequently, it is not a Mueller matrix.

b) $0 = \hat{d}_3 = \hat{d}_2 < \hat{d}_1$. $E_{\Delta d}$ degenerates into a segment. Therefore, $0 = l_{\Delta P3} = l_{\Delta P2} < l_{\Delta P1}$ and $l_{\Delta D3} = l_{\Delta D2} < l_{\Delta D1}$, so that $E_{\Delta P}$ and $E_{\Delta D}$ also degenerate into respective segments.

  b1) $\hat{d}_1 < 1$. The end-points of the segment $E_{\Delta d}$ do not touch the unit sphere. Consequently, the segments $E_{\Delta P}$ and $E_{\Delta D}$ (which, unlike $E_{\Delta d}$, are not located symmetrically with respect to the origin) do not touch the unit sphere either.

b2) $\hat{d}_1 = 1$. Both end-points of the segment $E_{\Delta d}$ touch the unit sphere. Consequently, both respective end-points of the segments $E_{\Delta P}$ and $E_{\Delta D}$ likewise touch the unit sphere.

c) $0 = \hat{d}_3 = \hat{d}_2 = \hat{d}_1$, $E_{\Delta d}$ degenerates into a single point located at the origin of the unit sphere. Therefore, $E_{\Delta P}$ and $E_{\Delta D}$ degenerate into respective single points located inside the unit sphere.

The above case analysis of the geometric features of Type-I singular depolarizers is summarized in Table 2.

Let us now consider type-II singular depolarizers. The determinant of a type-II Mueller matrix can be written as follows,

$$\det \mathbf{M} = \det \mathbf{M}_{J2} \det \mathbf{M}_{\Delta nd} \det \mathbf{M}_{J1}$$
$$= a_0^2 a_2^2 \det \mathbf{M}_{J2} \det \mathbf{M}_{J1}, \quad (36)$$

and, since pure Mueller matrices have nonnegative determinants, the determinant of a type-II Mueller matrix is nonnegative. Furthermore, for a type-II Mueller matrix, the inequalities $\det \mathbf{M}_{J1} > 0$ and $\det \mathbf{M}_{J2} > 0$ necessarily hold (otherwise $\mathbf{M}$ would correspond to a depolarizing analyzer or to a depolarizing polarizer, both being type-I matrices). Consequently, a type-II Mueller matrix is singular if and only if $a_2 = 0$ and furthermore, any singular type-II Mueller matrix is a type-II singular depolarizer.

The condition $a_2 = 0$ (while $a_0 > 0$ is assumed in order to avoid the trivial case of $\mathbf{M}_{S\Delta}$ being the zero matrix) entails that the canonical ellipsoid degenerates into a segment whose end-points have for coordinates $(1/3, 0, 0)$ and $(1, 0, 0)$. The forward and reverse ellipsoids of $\mathbf{M}_{S\Delta}$ likewise reduce to segments that touch the unit sphere at corresponding single points (given by the images of $(\pm 1, 0, 0)$ through $\mathbf{M}_{J2}$ and $\mathbf{M}_{J1}^T$ respectively).

The above case analysis of the geometric features of Type-II singular depolarizers is summarized in Table 3.

## 5. CONCLUSION

Through the framework of the forward [2], reverse [8] and symmetric [11] serial decompositions of depolarizing Mueller matrices, singular Mueller matrices have been interpreted and classified in a systematic way. Besides depolarizing polarizers $(P = 1, D < 1)$, depolarizing analyzers $(D = 1, P < 1)$ and polarizers $(D = P = 1)$, an additional family of depolarizing media, that of the *singular depolarizers*, simultaneously satisfying $P < 1$ and $D < 1$, features singular Mueller matrix.

Therefore, any medium satisfying $D = 1$ and/or $P = 1$ admits a serial decomposition including a polarizer and has a singular Mueller matrix. Furthermore, any medium satisfying $P < 1$ and $D < 1$ but whose symmetric decomposition includes a singular canonical depolarizing Mueller matrix, likewise has a singular Mueller matrix. The above-mentioned kinds of media cover all the families of singular Mueller matrices (depolarizing or not) and their properties are summarized in Tables 1, 2 and 3, which show their geometric representation through the corresponding characteristic ellipsoids:

- Polarizers are characterized by the fact that $E_{\Delta P}$ reduces to a single point lying on the surface of the unit sphere.
- Analyzers are characterized by the fact that $E_{\Delta D}$ reduces to a single point lying on the surface of the unit sphere.
- Pure polarizers are characterized by the fact that both $E_{\Delta P}$ and $E_{\Delta D}$ reduce to corresponding single points lying on the surface of the unit sphere.

- Type-II singular depolarizers are characterized by the fact that $E_\Delta$ (and hence, $E_{\Delta P}$ and $E_{\Delta D}$ too) reduces to a segment with one (and only one) of its end-points located on the surface of the unit sphere.

- type-I singular depolarizers are characterized by one of the following situations:
  - $E_{\Delta P}$ and $E_{\Delta D}$ are single points inside the unit sphere (but not on the surface). This occurs when $d_3 = d_2 = d_1 = 0$, so that $E_{\Delta d}$ is reduced to a single point located at the origin.
  - The three characteristic ellipsoids are segments. The fact that one of the characteristic ellipsoids is a segment implies that the other two have likewise the shapes of segments. The number of contact points with the surface of the unit sphere is the same for the three degenerate ellipsoids, with the following two possibilities:
    - Both end-points of each segment are located inside the unit sphere, without contact with its surface. This occurs when $0 = \hat{d}_3 = \hat{d}_2 < \hat{d}_1 < 1$.
    - Both end-points of each segment are located on the surface of the unit sphere. This occurs when $0 = \hat{d}_3 = \hat{d}_2$, $\hat{d}_1 = 1$.
  - The three characteristic ellipsoids degenerate into ellipses. The fact that one of the characteristic ellipsoids is an ellipse implies that the other two are likewise ellipses. The number of contact points with the surface of the unit sphere is the same for the three ellipses, with the following two possibilities:
    - The ellipses are located inside the unit sphere, without contact points with its surface. This occurs when $0 = \hat{d}_3 < \hat{d}_2 \leq \hat{d}_1 < 1$.
    - The ellipses have two contact points with the surface of the unit sphere. This occurs when $0 = \hat{d}_3 < \hat{d}_2 < \hat{d}_1 = 1$.

The authors believe that the above exhaustive classification of singular Mueller matrices will be useful to experimentalists willing to classify and interpret phenomenologically media with specific polarimetric responses.

**Funding Information.** Ministerio de Economía y Competitividad (FIS2011-22496 and FIS2014-58303-P); Gobierno de Aragón (E99).

## References

1. Z-F Xing, "On the deterministic and non-deterministic Mueller matrix," J. Mod Opt. **39**, 461-484 (1992).
2. S.-Y. Lu, R. A. Chipman, "Interpretation of Mueller matrices based on polar decomposition," J. Opt. Soc. Am. A **13**, 1106-1113 (1996).
3. J. J. Gil, "Polarimetric characterization of light and media," Eur. Phys. J. Appl. Phys. **40**, 1-47 (2007).
4. J. J. Gil, "Components of purity of a Mueller matrix," J. Opt. Soc. Am. A, **28**, 1578-1585 (2011).
5. Z. Sekera, "Scattering Matrices and Reciprocity Relationships for Various Representations of the State of Polarization," *J. Opt. Soc. Am.* **56**, 1732-1740 (1966).
6. A. Schönhofer and H. -G. Kuball, "Symmetry properties of the Mueller matrix," Chem. Phys. **115**, pp. 159-167 (1987).
7. J. J. Gil, E. Bernabéu, "Depolarization and polarization indices of an optical system" Opt. Acta **33**, 185-189 (1986).
8. J. Morio and F. Goudail, "Influence of the order of diattenuator, retarder and polarizer in polar decomposition of Mueller matrices," Opt. Lett. **29**, 2234-2236 (2004).

9. R. Ossikovski, A De Martino and S. Guyot, "Forward and reverse product decompositions of depolarizing Mueller matrices," Opt. Lett. **32**, 689-691 (2007).
10. J. J. Gil and I. San José, "Reduced form of a Mueller matrix," arXiv. 1509.06702 (2015).
11. R. Ossikovski, "Analysis of depolarizing Mueller matrices through a symmetric decomposition," J. Opt. Soc. Am. A **26**, 1109-1118 (2009).
12. C. V. M. van der Mee, "An eigenvalue criterion for matrices transforming Stokes parameters," J. Math. Phys. **34**, 5072-5088 (1993).
13. A. V. Gopala, K. S. Mallesh, Sudha, "On the algebraic characterization of a Mueller matrix in polarization optics. I Identifying a Mueller matrix from its *N* matrix," J. Mod. Optics **45**, 955-987 (1998).
14. A. V. Gopala, K. S. Mallesh, Sudha, "On the algebraic characterization of a Mueller matrix in polarization optics. II Necessary and sufficient conditions for Jones-derived Mueller matrices," J. Mod. Optics **45**, 989-999 (1998).
15. R. Ossikovski, "Canonical forms of depolarizing Mueller matrices," J. Opt. Soc. Am. A **27**, 123-130 (2010).
16. B. N. Simon, S. Simon, F. Gori, M. Santarsiero, R. Borghi, N. Mukunda and R. Simon, "Non-quantum entanglement resolves a basic issue in polarization optics", Phys. Rev. Lett. **104**, 023901-4 (2010).
17. B. N. Simon, S. Simon, N. Mukunda, F. Gori, M. Santarsiero, R. Borghi and R. Simon, "A complete characterization of pre-Mueller and Mueller matrices in polarization optics,". J. Opt. Soc. Am. A, **27**, 188-99 (2010).
18. J. J. Gil and R. Ossikovski, *Polarized light and the Mueller matrix approach*. 2016, CRC Press.
19. S.-Y. Lu and R. A. Chipman, "Mueller matrices and the degree of polarization," Opt. Commun. **146**, 11-14 (1998).
20. R. Ossikovski, J. J. Gil and I. San José, "Poincaré sphere mapping by Mueller matrices," J. Opt. Soc. Am. A **30**, 2291-2305 (2013).
21. J. J. Gil, E. Bernabéu, "A depolarization criterion in Mueller matrices," Opt. Acta **32**, 259-261 (1985).
22. J. J. Gil, "Review on Mueller matrix algebra for the analysis of polarimetric measurements," *J. Appl. Remote Sens.* **8**, 081599-37 (2014).
23. T. Tudor, "Generalized observables in polarization optics," J. Phys. A **36**, 9577-9590 (2003).
24. S.-Y. Lu and R. A. Chipman, "Homogeneous and inhomogeneous Jones matrices," J. Opt. Soc. Am. A **11**, 766-773 (1994).
25. N. G. Parke III, *Matrix optics*. PhD thesis, M.I.T. 1948.
26. J. J. Gil, *Determination of polarization parameters in matricial representation. Theoretical contribution and development of an automatic measurement device*. PhD thesis, University of Zaragoza, 1983. Available at http://zaguan.unizar.es/record/10680/files/TESIS-2013-057.pdf
27. S.R. Cloude, "Group theory and polarization algebra", Optik **75**, 26-36 (1986).
28. K. Kim, L. Mandel, E. Wolf "Relationship between Jones and Mueller matrices for random media" J. Opt. Soc. Am. A **4**, 433-437 (1987).

Table 1. Classification and properties of polarizers and analyzers.
*Points located on the surface of the unit sphere are represented by asterisks (∗), while points inside the unit sphere are represented by dots (.)*

| | Polarizance and diattenuation | Mueller matrix | Forward ellipsoid $E_{\Delta P}$ | Canonical ellipsoid $E_{\Delta}$ | Reverse ellipsoid $E_{\Delta D}$ |
|---|---|---|---|---|---|
| Depolarizing polarizer | $P=1$ $D<1$ | $\mathbf{M}_{\hat{P}D}$ $m_{00}\begin{pmatrix} 1 & \mathbf{D}^T \\ \hat{\mathbf{P}} & \hat{\mathbf{P}} \otimes \mathbf{D}^T \end{pmatrix}$ | ∗ (on surface) | . (inside) | . (inside) |
| Depolarizing analyzer | $P<1$ $D=1$ | $\mathbf{M}_{P\hat{D}}$ $m_{00}\begin{pmatrix} 1 & \hat{\mathbf{D}}^T \\ \mathbf{P} & \mathbf{P} \otimes \hat{\mathbf{D}}^T \end{pmatrix}$ | . (inside) | . (inside) | ∗ (on surface) |
| Polarizer | $P=1$ $D=1$ | $\mathbf{M}_{\hat{P}\hat{D}}$ $m_{00}\begin{pmatrix} 1 & \hat{\mathbf{D}}^T \\ \hat{\mathbf{P}} & \hat{\mathbf{P}} \otimes \hat{\mathbf{D}}^T \end{pmatrix}$ | ∗ (on surface) | Undetermined | ∗ (on surface) |

Table 2. Classification and properties of type-I singular depolarizers.
*Points located on the surface of the unit sphere are represented by asterisks (∗), while points inside the unit sphere are represented by dots (.)*

| | Polarizance and diattenuation | Canonical depolarizer | Forward ellipsoid $E_{\Delta P}$ | Canonical ellipsoid $E_{\Delta}$ | Reverse ellipsoid $E_{\Delta D}$ |
|---|---|---|---|---|---|
| Type-I Singular depolarizer | $P<1$ $D<1$ $\det \mathbf{M}_{\Delta d}=0$ | $\mathbf{M}_{\Delta d}$ $\begin{pmatrix} d_0 & 0 & 0 & 0 \\ 0 & d_1 & 0 & 0 \\ 0 & 0 & d_2 & 0 \\ 0 & 0 & 0 & 0 \end{pmatrix}$ $d_2 \leq d_1 < d_0$ | | | |
| | | $\mathbf{M}_{\Delta d}$ $\begin{pmatrix} d_0 & 0 & 0 & 0 \\ 0 & d_0 & 0 & 0 \\ 0 & 0 & d_2 & 0 \\ 0 & 0 & 0 & 0 \end{pmatrix}$ $d_2 < d_0$ | | | |
| | | $\mathbf{M}_{\Delta d}$ $\begin{pmatrix} d_0 & 0 & 0 & 0 \\ 0 & d_1 & 0 & 0 \\ 0 & 0 & 0 & 0 \\ 0 & 0 & 0 & 0 \end{pmatrix}$ $d_1 < d_0$ | | | |
| | | $\mathbf{M}_{\Delta d}$ $d_0 \begin{pmatrix} 1 & 0 & 0 & 0 \\ 0 & 1 & 0 & 0 \\ 0 & 0 & 0 & 0 \\ 0 & 0 & 0 & 0 \end{pmatrix}$ | | | |
| | | $\mathbf{M}_{\Delta 0}$ $\begin{pmatrix} d_0 & 0 & 0 & 0 \\ 0 & 0 & 0 & 0 \\ 0 & 0 & 0 & 0 \\ 0 & 0 & 0 & 0 \end{pmatrix}$ | | | |

Table 3. Classification and properties of type-II singular depolarizers.
*Points located on the surface of the unit sphere are represented by asterisks (∗), while points inside the unit sphere are represented by dots (.)*

| | Polarizance and diattenuation | Canonical depolarizer | Forward ellipsoid $E_{\Delta P}$ | Canonical ellipsoid $E_{\Delta}$ | Reverse ellipsoid $E_{\Delta D}$ |
|---|---|---|---|---|---|
| Type-II Singular depolarizer | $P<1$ $D<1$ $\det \mathbf{M}_{\Delta nd}=0$ | $\mathbf{M}_{\Delta nd}$ $m_{00}\begin{pmatrix} 1 & -1/2 & 0 & 0 \\ 1/2 & 0 & 0 & 0 \\ 0 & 0 & 0 & 0 \\ 0 & 0 & 0 & 0 \end{pmatrix}$ | | | |